# Data Cube: A Relational Aggregation Operator Generalizing Group-By, Cross-Tab, and Sub-Totals

Jim Gray
Surajit Chaudhuri
Adam Bosworth
Andrew Layman
Don Reichart
Murali Venkatrao
Frank Pellow
Hamid Pirahesh[1]

Microsoft Research
Microsoft Corporation
One Microsoft Way
Redmond, WA 98052

[1] IBM Research, 500 Harry Road, San Jose, CA. 95120

# Data Cube: A Relational Aggregation Operator

# Generalizing Group-By, Cross-Tab, and Sub-Totals

Jim Gray
Surajit Chaudhuri
Adam Bosworth
Andrew Layman
Don Reichart
Murali Venkatrao
Frank Pellow[1]
Hamid Pirahesh[2]

Microsoft Research
Advanced Technology Division
Microsoft Corporation
One Microsoft Way
Redmond, WA 98052

[2] IBM Research, 500 Harry Road, San Jose, CA. 95120

# Data Cube: A Relational Aggregation Operator Generalizing Group-By, Cross-Tab, and Sub-Totals[3]

Jim Gray Microsoft Gray@Microsoft.com
Surajit Chaudhuri Microsoft SurajitC@Microsoft.com
Adam Bosworth Microsoft AdamB@Microsoft.com
Andrew Layman Microsoft AndrewL@Microsoft.com
Don Reichart Microsoft DonRei@Microsoft.com
Murali Venkatrao Microsoft MuraliV@Microsoft.com
Hamid Pirahesh IBM Pirahesh@Almaden.IBM.com
Frank Pellow IBM Pellow@vnet.IBM.com

***Abstract**: Data analysis applications typically aggregate data across many dimensions looking for anomalies or unusual patterns. The SQL aggregate functions and the* `GROUP BY` *operator produce zero-dimensional or one-dimensional aggregates. Applications need the N-dimensional generalization of these operators. This paper defines that operator, called the* **data cube** *or simply* **cube**. *The cube operator generalizes the histogram, cross-tabulation, roll-up, drill-down, and sub-total constructs found in most report writers. The novelty is that cubes are relations. Consequently, the cube operator can be imbedded in more complex non-procedural data analysis programs. The cube operator treats each of the N aggregation attributes as a dimension of N-space. The aggregate of a particular set of attribute values is a point in this space. The set of points forms an N-dimensional cube. Super-aggregates are computed by aggregating the N-cube to lower dimensional spaces. This paper (1) explains the cube and roll-up operators, (2) shows how they fit in SQL, (3) explains how users can define new aggregate functions for cubes, and (4) discusses efficient techniques to compute the cube. Many of these features are being added to the SQL Standard.*

## 1. Introduction

Data analysis applications look for unusual patterns in data. They categorize data values and trends, extract statistical information, and then contrast one category with another. There are four steps to such data analysis:

- **formulating** a query that extracts relevant data from a large database,
- **extracting** the aggregated data from the database into a file or table,
- **visualizing** the results in a graphical way, and
- **analyzing** the results and formulating a new query.

Visualization tools display data trends, clusters, and differences. Some of the most exciting work in visualization focuses on presenting new graphical metaphors that allow people to discover data trends and anomalies. Many of these visualization and data analysis tools represent the dataset as an *N*-dimensional space. Visualization tools render two and three-dimensional sub-slabs of this space as 2D or 3D objects.

Color and time (motion) add two more dimensions to the display giving the potential for a 5D display. A Spreadsheet application such as Excel is an example of a data visualization/analysis tool that is used widely. Data analysis tools often try to identify a subspace of the N-dimensional space which is "interesting" (e.g., discriminating attributes of the data set).

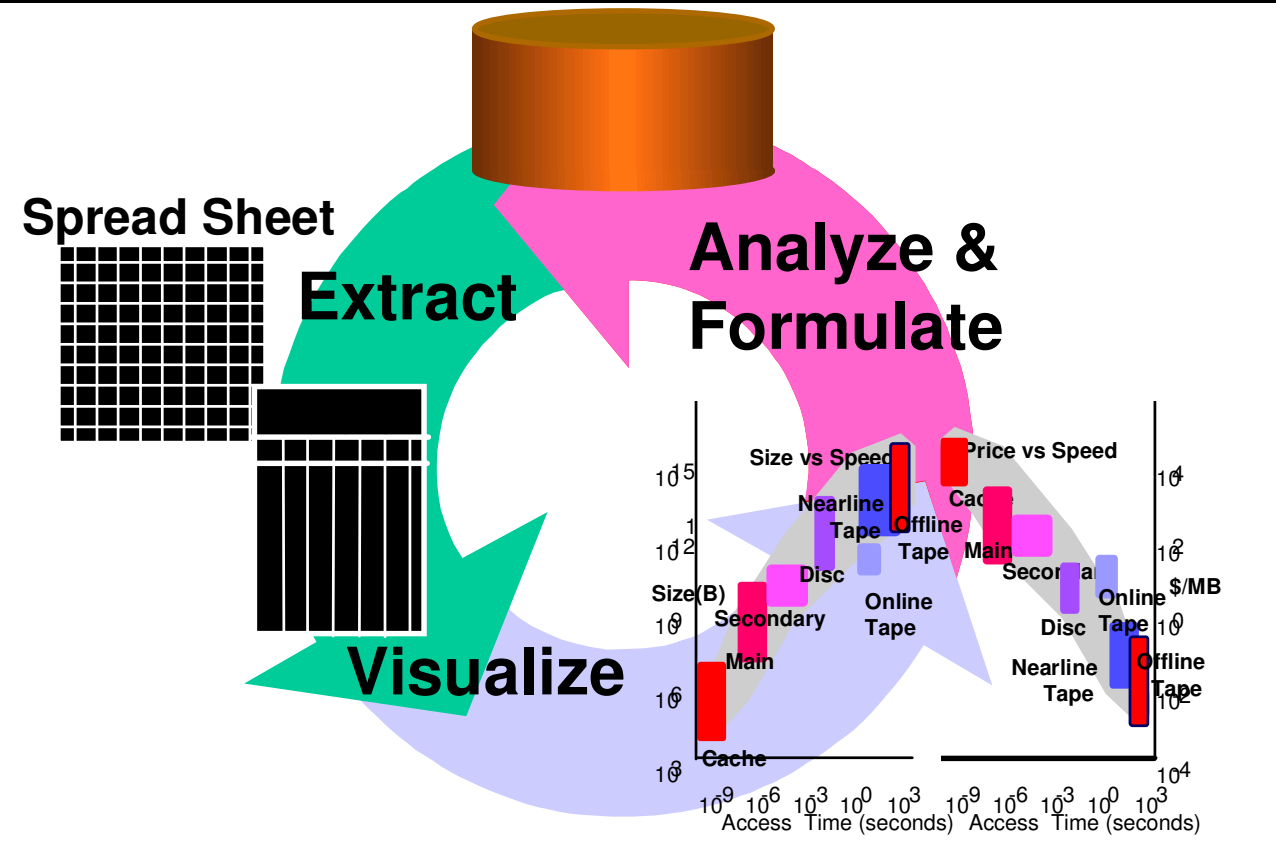

**Figure 1:** Data analysis tools facilitate the Extract-Visualize-Analyze loop. The cube and roll-up operators along with system and user-defined aggregates are part of the extraction process.

Thus, visualization as well as data analysis tools do "dimensionality reduction" , often by summarizing data along the dimensions that are left out. For example, in trying to analyze car sales, we might focus on the role of model, year and color of the cars in sale. Thus, we ignore the differences between two sales along the dimensions of date of sale or dealership but analyze the totals sale for cars by model, by year and by color only. Along with summarization and dimensionality reduction, data analysis applica-tions use constructs such as histogram, cross-tabulation, subtotals, roll-up and drill-down extensively.

[3] An extended abstract of this paper appeared in [Gray et.al.]

This paper examines how a relational engine can support efficient extraction of information from a SQL database that matches the above requirements of the visualization and data analysis. We begin by discussing the relevant features in Standard SQL and some of the vendor-specific SQL extensions. Section 2 discusses why `GROUP BY` fails to adequately address the requirements. The Cube and the ROLLUP operators are introduced in Section 3 and we also discuss how these operators overcome some of the shortcomings of `GROUP BY`. Sections 4 and 5 discuss how we can address and compute the Cube.

## 1.1. Relational and SQL Data Extraction

How do traditional relational databases fit into this multi-dimensional data analysis picture? How can 2D flat files (SQL tables) model an *N*-dimensional problem? Furthermore, how do the relational systems support the ability to support operations over N-dimensional representation that are central to visualization and data analysis programs?
We address each of these two issues in this section. The answer to the first question is that relational systems model *N*-dimensional data as a relation with *N*-attribute domains. For example, 4-dimensional (4D) earth temperature data is typically represented by a `Weather` table (Table 1). The first four columns represent the four dimensions: latitude, longitude, altitude, and time. Additional columns represent measurements at the 4D points such as temperature, pressure, humidity, and wind velocity. Each individual weather measurement is recorded as a new row of this table. Often these measured values are aggregates over time (the hour) or space (a measurement area centered on the point).

**Table 1:** `Weather`

| Time (UCT) | Latitude | Longitude | Altitude (m) | Temp (c) | Pres (mb) |
|---|---|---|---|---|---|
| 96/6/1:1500 | 37:58:33N | 122:45:28W | 102 | 21 | 1009 |
| many more rows like the ones above and below | | | | | |
| 96/6/7:1500 | 34:16:18N | 27:05:55W | 10 | 23 | 1024 |

As mentioned in the introduction, visualization and data analysis tools extensively use dimensionality reduction (aggregation) for better comprehensibility. Often data along the other dimensions that are not included in a "2-D" representation are summarized via aggregation in the form of histogram, cross-tabulation, subtotals etc. In SQL Standard, we depend on aggregate functions and the Group By operator to support aggregation.

The SQL standard [SQL], [Melton, Simon] provides five functions to aggregate the values in a table: `COUNT()`, `SUM()`, `MIN()`, `MAX()`, and `AVG()`. For example, the average of all measured temperatures is expressed as:

```
SELECT  AVG(Temp)
FROM    Weather;
```

In addition, SQL allows aggregation over distinct values. The following query counts the distinct number of reporting times in the `Weather` table:

```
SELECT  COUNT(DISTINCT Time)
FROM    Weather;
```

Aggregate functions return a single value. Using the `GROUP BY` construct, SQL can also create a table of many aggregate values indexed by a set of attributes. For example, The following query reports the average temperature for each reporting time and altitude:

```
SELECT   Time, Altitude, AVG(Temp)
FROM     Weather
GROUP BY Time, Altitude;
```

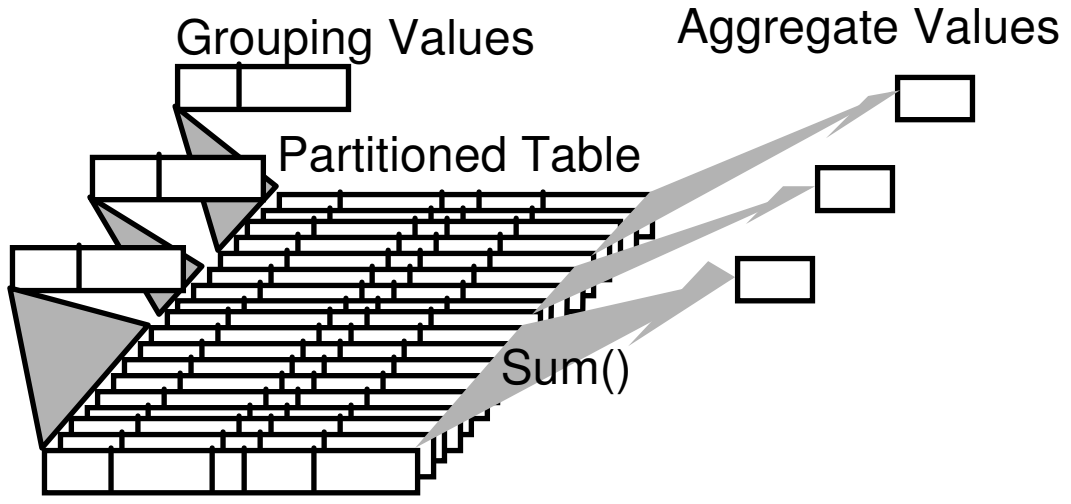

**Figure 2**: The `GROUP BY` relational operator partitions a table into groups. Each group is then aggregated by a function. The aggregation function summarizes some column of groups returning a value for each group.

`GROUP BY` is an unusual relational operator: It partitions the relation into disjoint tuple sets and then aggregates over each set as illustrated in Figure 2.

SQL's aggregation functions are widely used in database applications.Thispopularity is reflected in the presence of a large number of queries in the decision-support benchmark TPC-D [TPC]. The TPC-D query set has one 6D `GROUP BY` and three 3D `GROUP BY`s. One and two dimensional `GROUP BY`s are the most common. Surprisingly, aggregates appear in the TPC online-transaction processing benchmarks as well (TPC-A, B and C) . Table 2 shows how frequently the database and transaction processing benchmarks use aggregation and `GROUP BY`. A detailed description of these benchmarks is beyond the scope of the paper (See [Gray] and [TPC]).

| Table 2: SQL Aggregates in Standard Benchmarks. | | | |
|---|---|---|---|
| Benchmark | Queries | Aggregates | GROUP BYs |
| TPC-A, B | 1 | 0 | 0 |
| TPC-C | 18 | 4 | 0 |
| TPC-D | 16 | 27 | 15 |
| Wisconsin | 18 | 3 | 2 |
| $AS^3AP$ | 23 | 20 | 2 |
| SetQuery | 7 | 5 | 1 |

## 1.2. Extensions In Some SQL Systems

Beyond the five standard aggregate functions defined so far, many SQL systems add statistical functions (median, standard deviation, variance, etc.), physical functions (center of mass, angular momentum, etc.), financial analysis (volatility, Alpha, Beta, etc.), and other domain-specific functions.

Some systems allow users to add new aggregation functions. The Informix Illustra system, for example, allows users to add aggregate functions by adding a program with the following three callbacks to the database system [Informix]:

**`Init`** `(&handle)`: Allocates the handle and initializes the aggregate computation.

**`Iter`** `(&handle, value)`: Aggregates the next value into the current aggregate.

`value =` **`Final`**`(&handle)`: Computes and returns the resulting aggregate by using data saved in the handle. This invocation deallocates the handle.

Consider implementing the `Average()` function. The `handle` stores the `count` and the `sum` initialized to `zero`. When passed a new non-null value, `Iter()` increments the `count` and adds the `sum` to the value. The `Final()` call deallocates the `handle` and returns `sum` divided by `count`. IBM's DB2 Common Server [Chamberlin] has a similar mechanism. This design has been added to the Draft Proposed standard for SQL.[SQL97].

Red Brick systems, one of the larger UNIX OLAP vendors, add some interesting aggregate functions that enhance the `GROUP BY` mechanism [Red Brick]:

**`Rank`**`(expression)`: returns the expression's rank in the set of all values of this domain of the table. If there are *N* values in the column, and this is the highest value, the rank is *N*, if it is the lowest value the rank is 1.

**`N_tile`**`(expression, n)`: The range of the expression (over all the input values of the table) is computed and divided into `n` value ranges of approximately equal population. The function returns the number of the range containing the expression's value. If your bank account was among the largest 10% then your `rank(account.balance,10)` would return 10. Red Brick provides just `N_tile(expression,3)`.

**`Ratio_To_Total`**`(expression)`: Sums all the expressions. Then for each instance, divides the expression instance by the total sum.

To give an example, the following SQL statement

```
SELECT  Percentile, MIN(Temp), MAX(Temp)
FROM    Weather
GROUP BY N_tile(Temp,10) as Percentile
HAVING  Percentile = 5;
```

returns one row giving the minimum and maximum temperatures of the middle 10% of all temperatures.

Red Brick also offers three **cumulative aggregates** that operate on ordered tables.

**`Cumulative`**`(expression)`: Sums all values so far in an ordered list.

**`Running_Sum`**`(expression,n)`: Sums the most recent `n` values in an ordered list. The initial `n-1` values are `NULL`.

**`Running_Average`**`(expression,n)`: Averages the most recent `n` values in an ordered list. The initial `n-1` values are `NULL`.

These aggregate functions are optionally reset each time a grouping value changes in an ordered selection.

# 2. Problems With `GROUP BY`:

Certain common forms of data analysis are difficult with these SQL aggregation constructs. As explained next, three common problems are: (1) Histograms, (2) Roll-up Totals and Sub-Totals for drill-downs, (3) Cross Tabulations.

The standard SQL `GROUP BY` operator does not allow a direct construction of **histograms** (aggregation over computed categories). For example, for queries based on the `Weather` table, it would be nice to be able to group times into days, weeks, or months, and to group locations into areas (e.g., US, Canada, Europe,...). If a `Nation()` function maps latitude and longitude into the name of the country containing that location, then the following query would give the daily maximum reported temperature for each nation.

```
SELECT    day, nation, MAX(Temp)
FROM      Weather
GROUP BY  Day(Time) AS day,
          Nation(Latitude , Longitude)
                AS nation;
```

Some SQL systems support histograms directly but the standard does not[4]. In standard SQL, histograms are computed indirectly from a table-valued expression which is then aggregated. The following statement demonstrates this SQL92 construct using nested queries.

[4] These criticisms led to a proposal to include theses features in the draft SQL standard [SQL97].

```
SELECT day, nation, MAX(Temp)
FROM ( SELECT Day(Time) AS day,
              Nation(Latitude, Longitude)
                                AS nation,
              Temp
       FROM Weather
     ) AS foo
GROUP BY day, nation;
```

A more serious problem, and the main focus of this paper, relates to roll-ups using totals and **sub-totals** for **drill-down** reports. Reports commonly aggregate data at a coarse level, and then at successively finer levels. The car sales report in Table 3 shows the idea (this and other examples are based on the sales summary data in the table in Figure 4) . Data is aggregated by Model, then by Year, then by Color. The report shows data aggregated at three levels. Going up the levels is called **rolling-up** the data. Going down is called **drilling-down** into the data. Data aggregated at each distinct level produces a sub-total.

**Table 3.a:** Sales Roll Up by Model by Year by Color

| **Model** | **Year** | **Color** | **Sales by Model by Year by Color** | **Sales by Model by Year** | **Sales by Model** |
|---|---|---|---|---|---|
| Chevy | 1994 | black | 50 | | |
| | | white | 40 | | |
| | | | | 90 | |
| | 1995 | black | 85 | | |
| | | white | 115 | | |
| | | | | 200 | |
| | | | | | 290 |

Table 3.a suggests creating $2^N$ aggregation columns for a roll-up of *N* elements. Indeed, Chris Date recommends this approach [Date1]. His design gives rise to Table 3.b

**Table 3.b:** Sales Roll-Up by Model by Year by Color as recommended by Chris Date [Date1].

| **Model** | **Year** | **Color** | **Sales** | **Sales by Model by Year** | **Sales by Model** |
|---|---|---|---|---|---|
| Chevy | 1994 | black | 50 | 90 | 290 |
| Chevy | 1994 | white | 40 | 90 | 290 |
| Chevy | 1995 | black | 85 | 200 | 290 |
| Chevy | 1995 | white | 115 | 200 | 290 |

The representation of Table 3.a is not relational because the empty cells (presumably `NULL` values), cannot form a key. Representation 3.b is an elegant solution to this problem, but we rejected it because it implies enormous numbers of domains in the resulting tables. We were intimidated by the prospect of adding 64 columns to the answer set of a 6D TPCD query. The representation of Table 3.b is also not convenient -- the number of columns grows as the power set of the number of aggregated attributes, creating difficult naming problems and very long names. The approach recommended by Date is reminiscent of pivot tables found in Excel (and now all other spreadsheets) [Excel], a popular data analysis feature of Excel[5].

**Table 4:** An Excel pivot table representation of Table 3 with Ford sales data included.

| Sum | Year | Color | | | | | |
|---|---|---|---|---|---|---|---|
| Sales | 1994 | | 1994 Total | 1995 | | 1995 Total | Grand Total |
| Model | black | white | | black | white | | |
| Chevy | 50 | 40 | 90 | 85 | 115 | 200 | 290 |
| Ford | 50 | 10 | 60 | 85 | 75 | 160 | 220 |
| Grand Total | 100 | 50 | 150 | 170 | 190 | 360 | 510 |

Table 4 an alternative representation of Table 3a (with Ford Sales data included) that illustrates how a pivot table in Excel can present the Sales data by Model, by Year, and then by Color. The pivot operator transposes a spreadsheet: typically aggregating cells based on values in the cells. Rather than just creating columns based on subsets of column names, pivot creates columns based on subsets of column *values*. This is a *much* larger set If one pivots on two columns containing *N* and *M* values, the resulting pivot table has *NxM* values. We cringe at the prospect of so many columns and such obtuse column names.

Rather than extend the result table to have many new columns, a more conservative approach prevents the exponential growth of columns by overloading column values. The idea is to introduce an ALL value. Table 5.a demonstrates this relational and more convenient representation. The dummy value "`ALL`" has been added to fill in the super-aggregation items.:

**Table 5.a**: Sales Summary

| **Model** | **Year** | **Color** | **Units** |
|---|---|---|---|
| Chevy | 1994 | black | 50 |
| Chevy | 1994 | white | 40 |
| Chevy | 1994 | `ALL` | 90 |
| Chevy | 1995 | black | 85 |
| Chevy | 1995 | white | 115 |
| Chevy | 1995 | `ALL` | 200 |
| Chevy | `ALL` | `ALL` | 290 |

5 It seems likely that a relational pivot operator will appear in database systems in the near future.

Table 5.a is not really a completely new representation or operation. Since Table 5.a is a relation, it is not surprising that it can be built using standard SQL. The SQL statement to build this `SalesSummary` table from the raw `Sales` data is:

```
SELECT ‘ALL’, ‘ALL’, ‘ALL’, SUM(Sales)
   FROM      Sales
   WHERE     Model = 'Chevy'
UNION
SELECT Model, ‘ALL’, ‘ALL’, SUM(Sales)
   FROM      Sales
   WHERE     Model = 'Chevy'
  GROUP BY   Model
UNION
SELECT Model, Year, ‘ALL’, SUM(Sales)
   FROM      Sales
   WHERE     Model = 'Chevy'
   GROUP BY  Model, Year
UNION
SELECT Model, Year, Color, SUM(Sales)
   FROM      Sales
   WHERE     Model = 'Chevy'
   GROUP BY  Model, Year, Color;
```

This is a simple 3-dimensional roll-up. Aggregating over *N* dimensions requires *N* such unions.

Roll-up is asymmetric – notice that Table 5.a aggregates sales by year but not by color. These rows are:

**Table 5.b**: Sales Summary rows missing form Table 5.a to convert the roll-up into a cube.

| Model | Year | Color | Units |
|---|---|---|---|
| Chevy | ALL | black | 135 |
| Chevy | ALL | white | 155 |

These additional rows could be captured by adding the following clause to the SQL statement above:

```
UNION
SELECT Model, ‘ALL’, Color, SUM(Sales)
   FROM      Sales
   WHERE     Model = 'Chevy'
   GROUP BY  Model, Color;
```

The symmetric aggregation result is a table called a **cross-tabulation,** or **cross tab** for short. Tables 5.a and 5.b are the relational form of the cross-tabs, but cross tab data is routinely displayed in the more compact format of Table 6.

**Table 6.a: Chevy Sales Cross Tab**

| **Chevy** | **1994** | **1995** | **total** (ALL) |
|---|---|---|---|
| **black** | 50 | 85 | 135 |
| **white** | 40 | 115 | 155 |
| **total** (ALL) | 90 | 200 | 290 |

This cross tab is a two-dimensional aggregation. If other automobile models are added, it becomes a 3D aggregation. For example, data for Ford products adds an additional cross tab plane.

The cross-tab-array representation (Table 6.a, 6.b) is equivalent to the relational representation using the `ALL` value. Both generalize to an *N*-dimensional cross tab. Most report writers build in a cross-tabs feature, building the report up from the underlying tabular data such as Table 5. See for example the `TRANSFORM-PIVOT` operator of Microsoft Access [Access].

**Table 6b: Ford Sales Cross Tab**

| **Ford** | **1994** | **1995** | **total** (ALL) |
|---|---|---|---|
| **black** | 50 | 85 | 135 |
| **white** | 10 | 75 | 85 |
| **total** (ALL) | 60 | 160 | 220 |

The representation suggested by Table 5 and unioned `GROUP BY`s “solve” the problem of representing aggregate data in a relational data model. The problem remains that expressing roll-up, and cross-tab queries with conventional SQL is daunting. A six dimension cross-tab requires a 64-way union of 64 different `GROUP BY` operators to build the underlying representation.

There is another very important reason why it is inadequate to use GROUP BYs. The resulting representation of aggregation is too complex to analyze for optimization. On most SQL systems this will result in 64 scans of the data, 64 sorts or hashes, and a long wait.

# 3. CUBE and ROLLUP Operators

The generalization of group by, roll-up and cross-tab ideas seems obvious: Figure 3 shows the concept for aggregation up to 3-dimensions. The traditional `GROUP BY` generates the *N*-dimensional data cube *core*. The *N-1* lower-dimensional aggregates appear as points, lines, planes, cubes, or hyper-cubes hanging off the data cube core.

The data cube operator builds a table containing all these aggregate values. The total aggregate using function `f()` is represented as the tuple:

```
  ALL, ALL, ALL,..., ALL, f(*)
```

Points in higher dimensional planes or cubes have fewer `ALL` values.

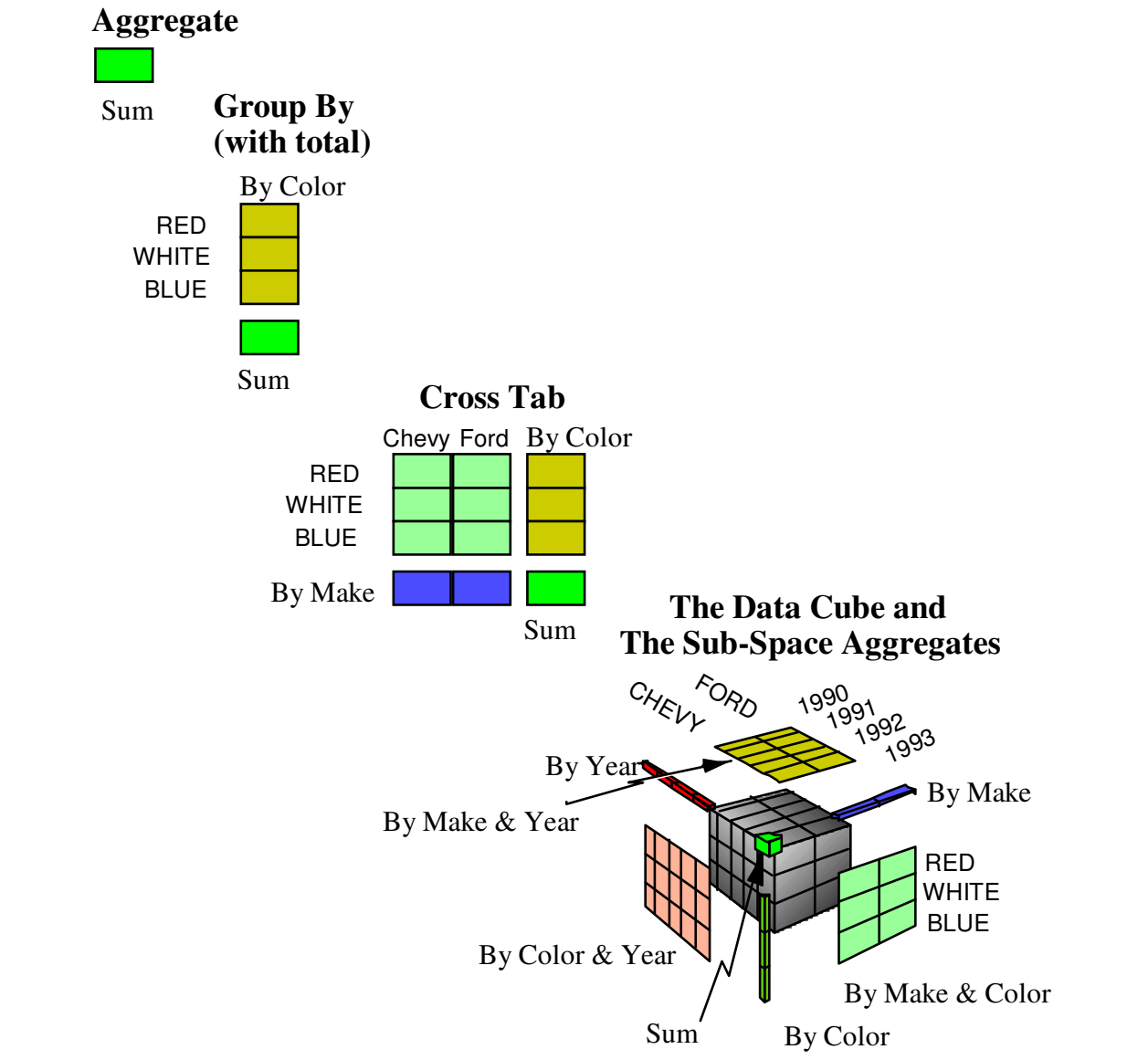

**Figure 3:** The CUBE operator is the *N*-dimensional generalization of simple aggregate functions. The 0D data cube is a point. The 1D data cube is a line with a point. The 2D data cube is a cross tabulation, a plane, two lines, and a point. The 3D data cube is a cube with three intersecting 2D cross tabs.

Creating a data cube requires generating the power set (set of all subsets) of the aggregation columns. Since the CUBE is an aggregation operation, it makes sense to externalize it by overloading the SQL GROUP BY operator. In fact, the CUBE is a relational operator, with GROUP BY and ROLL UP as degenerate forms of the operator. This can be conveniently specified by overloading the `SQL GROUP BY`[6].

Figure 4 has an example of the cube syntax. To give another, here follows a statement to aggregate the set of temperature observations:

```
SELECT    day, nation, MAX(Temp)
FROM      Weather
GROUP BY CUBE
          Day(Time) AS day,
          Country(Latitude, Longitude)
                                AS nation;
```

The semantics of the `CUBE` operator are that it first aggregates over all the `<select list>` attributes in the `GROUP BY` clause as in a standard `GROUP BY`. Then, it `UNIONs` in each super-aggregate of the global cube -- substituting `ALL` for the aggregation columns. If there are *N* attributes in the `<select list>`, there will be $2^N-1$ super-aggregate values. If the cardinality of the *N* attributes are $C_1, C_2, \ldots, C_N$ then the cardinality of the resulting cube relation is $\prod(C_i + 1)$. The extra value in each domain is `ALL`. For example, the SALES table has 2x3x3 = 18 rows, while the derived data cube has 3x4x4 = 48 rows.

```
SELECT Model, Year, Color, SUM(sales) AS Sales
FROM Sales
WHERE Model in {'Ford', 'Chevy'}
  AND Year BETWEEN 1990 AND 1992
GROUP BY CUBE Model, Year, Color;
```

SALES

| Model | Year | Color | Sales |
|---|---|---|---|
| Chevy | 1990 | red | 5 |
| Chevy | 1990 | white | 87 |
| Chevy | 1990 | blue | 62 |
| Chevy | 1991 | red | 54 |
| Chevy | 1991 | white | 95 |
| Chevy | 1991 | blue | 49 |
| Chevy | 1992 | red | 31 |
| Chevy | 1992 | white | 54 |
| Chevy | 1992 | blue | 71 |
| Ford | 1990 | red | 64 |
| Ford | 1990 | white | 62 |
| Ford | 1990 | blue | 63 |
| Ford | 1991 | red | 52 |
| Ford | 1991 | white | 9 |
| Ford | 1991 | blue | 55 |
| Ford | 1992 | red | 27 |
| Ford | 1992 | white | 62 |
| Ford | 1992 | blue | 39 |

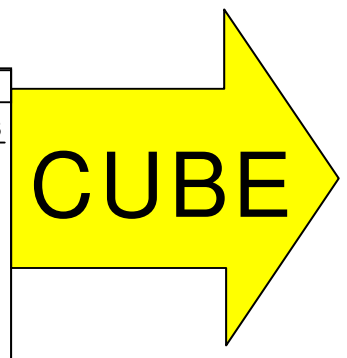

DATA CUBE

| Model | Year | Color | Sales |
|---|---|---|---|
| Chevy | 1990 | blue | 62 |
| Chevy | 1990 | red | 5 |
| Chevy | 1990 | white | 95 |
| Chevy | 1990 | ALL | 154 |
| Chevy | 1991 | blue | 49 |
| Chevy | 1991 | red | 54 |
| Chevy | 1991 | white | 95 |
| Chevy | 1991 | ALL | 198 |
| Chevy | 1992 | blue | 71 |
| Chevy | 1992 | red | 31 |
| Chevy | 1992 | white | 54 |
| Chevy | 1992 | ALL | 156 |
| Chevy | ALL | blue | 182 |
| Chevy | ALL | red | 90 |
| Chevy | ALL | white | 236 |
| Chevy | ALL | ALL | 508 |
| Ford | 1990 | blue | 63 |
| Ford | 1990 | red | 64 |
| Ford | 1990 | white | 62 |
| Ford | 1990 | ALL | 189 |
| Ford | 1991 | blue | 55 |
| Ford | 1991 | red | 52 |
| Ford | 1991 | white | 9 |
| Ford | 1991 | ALL | 116 |
| Ford | 1992 | blue | 39 |
| Ford | 1992 | red | 27 |
| Ford | 1992 | white | 62 |
| Ford | 1992 | ALL | 128 |
| Ford | ALL | blue | 157 |
| Ford | ALL | red | 143 |
| Ford | ALL | white | 133 |
| Ford | ALL | ALL | 433 |
| ALL | 1990 | blue | 125 |
| ALL | 1990 | red | 69 |
| ALL | 1990 | white | 149 |
| ALL | 1990 | ALL | 343 |
| ALL | 1991 | blue | 106 |
| ALL | 1991 | red | 104 |
| ALL | 1991 | white | 110 |
| ALL | 1991 | ALL | 314 |
| ALL | 1992 | blue | 110 |
| ALL | 1992 | red | 58 |
| ALL | 1992 | white | 116 |
| ALL | 1992 | ALL | 284 |
| ALL | ALL | blue | 339 |
| ALL | ALL | red | 233 |
| ALL | ALL | white | 369 |
| ALL | ALL | ALL | 941 |

**Figure 4:** A 3D data cube (right) built from the table at the left by the `CUBE` statement at the top of the figure.

If the application wants only a roll-up or drill-down report, similar to the data in Table 3.a, the full cube is overkill. Indeed, some parts of the full cube may be meaningless. If the answer set is not is not normalized, there may be functional dependencies among columns. For example, a date functionally defines a week, month, and year. Roll-ups by year, week, day are common, but a cube on these three attributes would be meaningless.

The solution is to offer `ROLLUP` in addition to `CUBE`. `ROLLUP` produces just the super-aggregates:

```
(v1 ,v2 ,...,vn,  f()),
(v1 ,v2 ,...,ALL, f()),

          ...
(v1 ,ALL,...,ALL, f()),
(ALL,ALL,...,ALL, f()).
```

Cumulative aggregates, like running sum or running average, work especially well with `ROLLUP` because the answer set is naturally sequential (linear) while the `full data cube` is naturally non-linear (multi-dimensional). `ROLLUP` and `CUBE` must be ordered for cumulative operators to apply.

We investigated letting the programmer specify the exact list of super-aggregates but encountered complexities related to collation, correlation, and expressions. We believe `ROLLUP` and `CUBE` will serve the needs of most applications.

[6] An earlier version of this paper [Gray et. al.] and the Microsoft SQL Server 6.5 product implemented a slightly different syntax. They suffix the `GROUP BY` clause with a `ROLLUP` or `CUBE` modifier. The SQL Standards body chose an infix notation so that `GROUP BY` and `ROLLUP` and `CUBE` could be mixed in a single statement. The improved syntax is described here.

## 3.1. The GROUP, CUBE, ROLLUP Algebra

The GROUP BY, ROLLUP, and CUBE operators have an interesting algebra. The CUBE of a ROLLUP or GROUP BY is a CUBE. The ROLLUP of a GROUP BY is a ROLLUP. Algebraically, this operator algebra can be stated as:

```
        CUBE(ROLLUP) = CUBE
        ROLLUP(GROUP BY) = ROLLUP
```

So it makes sense to arrange the aggregation operators in the compound order where the "most powerful" cube operator at the core, then a roll-up of the cubes and then a group by of the roll-ups. Of course, one can use any subset of the three operators:

```
GROUP BY <select list>
        ROLLUP <select list>
                CUBE <select list>
```

The following SQL demonstrates a compound aggregate. The "shape" of the answer is diagrammed in Figure 5:

```
SELECT Manufacturer,
            Year , Month,  Day,
               Color, Model
                    SUM(price) AS Revenue
FROM    Sales
GROUP BY Manufacturer,
   ROLLUP   Year(Time) AS Year ,
            Month(Time) AS Month,
            Day(Time) AS Day,
        CUBE
              Color,
              Model;
```

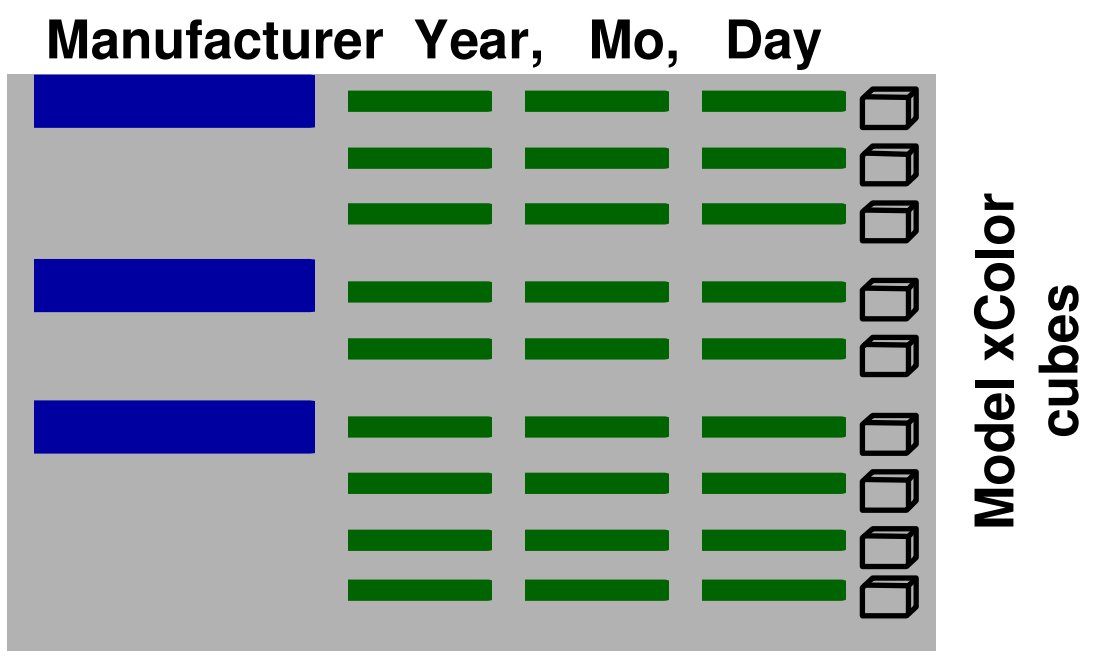

**Figure 5.** The combination of a GROUP BY on Manufacture, ROLLUP on year, month, day, and CUBE on some attributes. The aggregate values are the contents of the cube.

## 3.2. A Syntax Proposal

With these concepts in place, the syntactic extension to SQL is fairly easily defined. The current SQL GROUP BY syntax is:

```
GROUP BY
        {<column name> [collate clause]  ,...}
```

To support histograms and other function-valued aggregations, we first extend the GROUP BY syntax to:

```
GROUP BY <aggregation list>
<aggregation list> ::=
     { ( <column name> | <expression>)
              [ AS <correlation name>    ]
              [ <collate clause>         ]
                ,...}
```

These extensions are independent of the CUBE operator. They remedy some pre-existing problems with GROUP BY. Many systems already allow these extensions.

Now extend SQL's GROUP BY operator:

```
 GROUP BY [<aggregation list> ]
        [ ROLLUP <aggregation list> ]
            [ CUBE <aggregation list> ]
```

## 3.3. A Discussion of the ALL Value

Is the ALL value really needed? Each ALL value really represents a set – the set over which the aggregate was computed[7]. In the Table 5 SalesSummary data cube, the respective sets are:

```
    Model.ALL = ALL(Model)  = {Chevy, Ford }
    Year.ALL  = ALL(Year)   = {1990,1991,1992}
    Color.ALL = ALL(Color)  = {red,white,blue}
```

In reality, we have stumbled in to the world of nested relations – relations can be values. This is a major step for relational systems. There is much debate on how to proceed. Rather than attack those problems here, we just use the ALL value as a token representing these sets. Thinking of the ALL value as the corresponding set defines the semantics of the relational operators (e.g., equals and IN). The ALL string is for display. A new ALL() function generates the set associated with this value as in the examples above. ALL() applied to any other value returns NULL. This design may be eased by SQL3's support for set-valued variables and domains.

The ALL value appears to be essential, but creates substantial complexity. It is a non-value, like NULL. We do not add it lightly – adding it touches many aspects of the SQL language. To name a few:

- Treating each ALL value as the set of aggregates guides the meaning of the ALL value.
- ALL becomes a new keyword denoting the set value.
- ALL [NOT] ALLOWED is added to the column definition syntax and to the column attributes in the system catalogs.
- ALL, like NULL, does not participate in any aggregate except COUNT().
- The set interpretation guides the meaning of the relational operators {=, <, <=, =, >=, >, IN}.

There are more such rules, but this gives a hint of the added complexity. As an aside, to be consistent, if the ALL value is a set then the other values of that domain must be treated as singleton sets in order to have uniform operators on the domain.

[7] This is distinct from saying that ALL represents *one* of the members of the set.

It is convenient to know when a column value is an aggregate. One way to test this is to apply the `ALL()` function to the value and test for a non-`NULL` value. This is so useful that we propose a Boolean function `GROUPING()` that, given a select list element, returns `TRUE` if the element is an `ALL` value, and `FALSE` otherwise.

## 3.4. Avoiding the ALL Value

Veteran SQL implementers will be terrified of the ALL value – like NULL, it will create many special cases. If the goal is to help report writer and GUI visualization software, then it may be simpler to adopt the following approach[8]:

- Use the `NULL` value in place of the `ALL` value.
- Do not implement the `ALL()` function.
- Implement the `GROUPING()` function to discriminate between `NULL` and `ALL`.

In this minimalist design, tools and users can simulate the `ALL` value as by for example:

```
SELECT  Model,Year,Color,SUM(sales),
                       GROUPING(Model),
                       GROUPING(Year),
                       GROUPING(Color)
FROM Sales
GROUP BY CUBE Model, Year, Color;
```

Wherever the `ALL` value appeared before, now the corresponding value will be `NULL` in the data field and `TRUE` in the corresponding grouping field. For example, the global sum of Figure 4 will be the tuple:

`(NULL,NULL,NULL,941,TRUE,TRUE,TRUE)`

rather than the tuple one would get with the "real" cube operator:

`( ALL,  ALL,  ALL,  941 ).`

## 3.5. Decorations

The next step is to allow *decorations,* columns that do not appear in the `GROUP BY` but that are functionally dependent on the grouping columns. Consider the example:

```
SELECT   department.name, sum(sales)
FROM     sales JOIN department
            USING (department_number)
GROUP BY sales.department_number;
```

The `department.name` column in the answer set is not allowed in current SQL, since it is neither an aggregation column (appearing in the `GROUP BY` list) nor is it an aggregate. It is just there to decorate the answer set with the name of the department. We recommend the rule that *if a decoration* column (or column value) is functionally dependent on the aggregation columns, then it may be included in the `SELECT` answer list.

[8] This is the syntax and approach used by Microsoft's SQL Server (version 6.5).

Decoration's interact with aggregate values. If the aggregate tuple functionally defines the decoration value, then the value appears in the resulting tuple. Otherwise the decoration field is `NULL`. For example, in the following query the `continent` is not specified unless nation is.

```
SELECT  day,nation,MAX(Temp),
          continent(nation) AS continent
FROM    Weather
GROUP BY CUBE
          Day(Time) AS day,
          Country(Latitude, Longitude)
                            AS nation
```

The query would produce the sample tuples:

**Table 7:** Demonstrating decorations and ALL

| **day** | **nation** | **max**(Temp) | **continent** |
|---|---|---|---|
| 25/1/1995 | USA | 28 | North America |
| ALL | USA | 37 | North America |
| 25/1/1995 | ALL | 41 | NULL |
| ALL | ALL | 48 | NULL |

## 3.6. Dimensions Star, and Snowflake Queries

While strictly not part of the `CUBE` and `ROLLUP` operator design, there is an important database design concept that facilitates the use of aggregation operations. It is common to record events and activities with a detailed record giving all the ***dimensions*** of the event. For example, the sales item record in Figure 6 gives the id of the buyer, seller, the product purchased, the units purchased, the price, the date and the sales office that is credited with the sale. There are probably many more dimensions about this sale, but this example gives the idea.

There are side tables that for each dimension value give its attributes. For example, the San Francisco sales office is in the Northern California District, the Western Region, and the US Geography. This fact would be stored in a dimension table for the Office[9]. The dimension table may also have decorations describing other attributes of that Office. These dimension tables define a spectrum of aggregation granularities for the dimension. Analysists might want to cube various dimensions and then aggregate or roll-up the cube up at any or all of these granularities.

[9] Database normalization rules [Date2] would recommend that the fact that the California District be stored once, rather than storing it once for each Office. So there might be an office, district, and region tables, rather than one big denormalize table. Query users find it convenient to use the denormalized table.

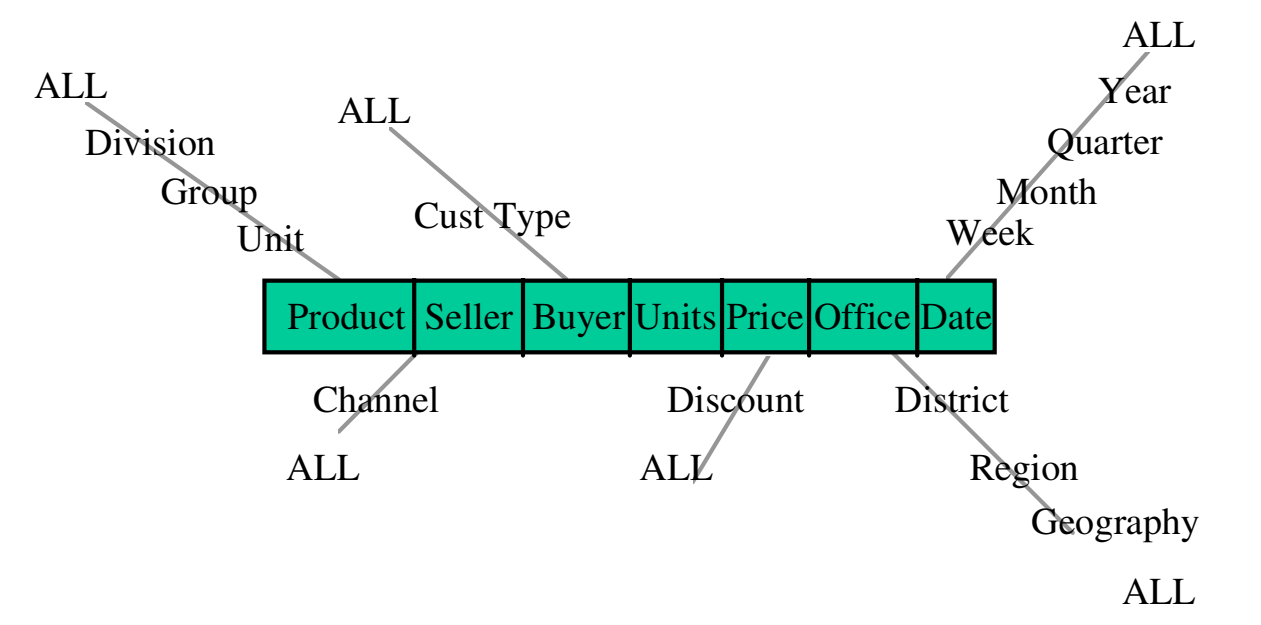

**Figure 6:** A snowflake schema showing the core fact table and some of the many aggregation ganularities of the core dimensions.

The general schema of Figure 6 is so common that it has been given a name: a **snowflake schema**. Simpler schemas that have a single dimension table for each dimension are called a **star schema**. Queries against these schemas are called **snowflake queries** and **star queries** respectively.

The diagram of Figure 6 suggests that the granularities form a pure hierarchy. In reality, the granularities typically form a lattice. To take just a very simple example, days nest in weeks but weeks do not nest in months or quarters or years (some weeks are partly in two years). Analysts often think of dates in terms of weekdays, weekends, sale days, various holidays (e.g., Christmas and the time leading up to it). So a fuller granularity graph of Figure 6 would be quite complex. Fortunately, graphical tools like pivot tables with pull down lists of categories hide much of this complexity from the analyst.

# 4. Addressing The Data Cube

Section 5 discusses how to compute data cubes and how users can add new aggregate operators. This section considers extensions to SQL syntax to easily access the elements of a data cube -- making it recursive and allowing aggregates to reference sub-aggregates.

It is not clear where to draw the line between the reporting-visualization tool and the query tool. Ideally, application designers should be able to decide how to split the function between the query system and the visualization tool. Given that perspective, the SQL system must be a Turing-complete programming environment.

SQL3 defines a Turing-complete procedural programming language. So, anything is possible. But, many things are not easy. Our task is to make simple and common things easy.

The most common request is for percent-of-total as an aggregate function. In SQL this is computed as a nested SELECT SQL statements.

```
SELECT Model,Year,Color,SUM(Sales),
       SUM(Sales)/
         (SELECT SUM(Sales)
         FROM Sales
        WHERE Model IN {'Ford','Chevy'}
        AND Year Between 1990 AND 1992
         )
FROM   Sales
WHERE  Model IN { 'Ford' , 'Chevy' }
  AND  Year Between 1990 AND 1992
GROUP BY CUBE Model, Year, Color ;
```

It seems natural to allow the shorthand syntax to name the global aggregate:

```
SELECT Model, Year, Color
                SUM(Sales) AS total,
     SUM(Sales) / total(ALL,ALL,ALL)
FROM Sales
WHERE Model IN { 'Ford' , 'Chevy' }
  AND Year Between 1990 AND 1992
GROUP BY CUBE Model, Year, Color;
```

This leads into deeper water. The next step is a desire to compute the *index* of a value -- an indication of how far the value is from the expected value. In a set of *N* values, one expects each item to contribute one *N*th to the sum. So the 1D index of a set of values is:

$$index(v_i) = v_i / (\Sigma_j v_j)$$

If the value set is two dimensional, this commonly used financial function is a nightmare of indices. It is best described in a programming language. The current approach to selecting a field value from a 2D `cube` would read as:

```
SELECT v
FROM    cube
WHERE   row    = :i
 AND    column = :j
```

We recommend the simpler syntax:

```
cube.v(:i, :j)
```

as a shorthand for the above selection expression. With this notation added to the SQL programming language, it should be fairly easy to compute super-super-aggregates from the base cube.

# 5. Computing Cubes and Roll-ups

`CUBE` and `ROLLUP` generalize aggregates and `GROUP BY`, so all the technology for computing those results also apply to computing the core of the cube [Graefe]. The basic technique for computing a `ROLLUP` is to sort the table on the aggregating attributes and then compute the aggregate functions (there is a more detailed discussion of the kind of aggregates in a moment.) If the `ROLLUP` result is small enough to fit in main memory, it can be computed by scanning the input set and applying each record to the in-memory `ROLLUP`. A cube is the union of many rollups, so the naive algorithm computes this union.

As Graefe [Graefe]. points out, the basic techniques for computing aggregates are:

- To minimize data movement and consequent processing cost, compute aggregates at the lowest possible system level.
- If possible, use arrays or hashing to organize the aggregation columns in memory, storing one aggregate value for each array or hash entry.
- If the aggregation values are large strings, it may be wise to keep a hashed symbol table that maps each string to an integer so that the aggregate values are small. When a new value appears, it is assigned a new integer. With this organization, the values become dense and the aggregates can be stored as an *N*-dimensional array.
- If the number of aggregates is too large to fit in memory, use sorting or hybrid hashing to organize the data by value and then aggregate with a sequential scan of the sorted data.
- If the source data spans many disks or nodes, use parallelism to aggregate each partition and then coalesce these aggregates.

Some innovation is needed to compute the "ALL" tuples of the cube and roll-up from the GROUP BY core. The ALL value adds one extra value to each dimension in the CUBE. So, an *N*-dimensional cube of *N* attributes each with cardinality $C_i$, will have $\prod(C_i+1)$. If each $C_i = 4$ then a 4D CUBE is 2.4 times larger than the base GROUP BY. We expect the $C_i$ to be large (tens or hundreds) so that the CUBE will be only a little larger than the GROUP BY. By comparison, an *N*-dimensional roll-up will add *only* N records to the answer set.

The cube operator allows many aggregate functions in the aggregation list of the GROUP BY clause. Assume in this discussion that there is a single aggregate function F() being computed on an *N*-dimensional cube. The extension to computing a list of functions is a simple generalization.

Figure 7 summarizes how aggregate functions are defined and implemented in many systems. It defines how the database execution engine initializes the aggregate function, calls the aggregate functions for each new value and then invokes the aggregate function to get the final value. More sophisticated systems allow the aggregate function to declare a computation cost so that the query optimizer knows to minimize calls to expensive functions. This design (except for the cost functions) is now part of the proposed SQL standard.

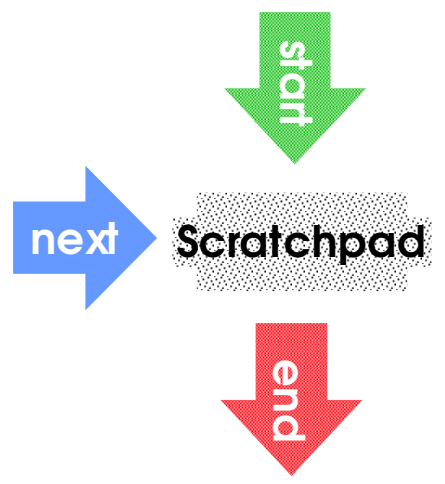

**Figure 7:** System defined and user defined aggregate functions are initialized with a start() call that allocates and initializes a scratchpad cell to compute the aggregate. Subsequently, the next() call is invoked for each value to be aggregated. Finally, the end() call computes the aggregate from the scratchpad values, deallocates the scratchpad and returns the result.

The simplest algorithm to compute the cube is to allocate a handle for each cube cell. When a new tuple: $(x_1, x_2, \ldots, x_N, v)$ arrives, the `Iter(handle, v)` function is called $2^N$ times -- once for each handle of each cell of the cube matching this value. The $2^N$ comes from the fact that each coordinate can either be $x_i$ or ALL. When all the input tuples have been computed, the system invokes the `final(&handle)` function for each of the $\prod(C_i+1)$ nodes in the cube. Call this the ***$2^N$-algorithm***. There is a corresponding order-*N* algorithm for roll-up.

If the base table has cardinality *T*, the *$2^N$-algorithm* invokes the `Iter()` function $T \times 2^N$ times. It is often faster to compute the super-aggregates from the core GROUP BY, reducing the number of calls by approximately a factor of *T*. It is often possible to compute the cube from the core or from intermediate results only *M* times larger than the core. The following trichotomy characterizes the options in computing super-aggregates.

Consider aggregating a two dimensional set of values $\{X_{ij} \mid i = 1,\ldots,I;\ j=1,\ldots,J\}$. Aggregate functions can be classified into three categories:

**Distributive**: Aggregate function *F()* is distributive if there is a function *G()* such that $F(\{X_{i,j}\}) = G(\{F(\{X_{i,j} \mid i=1,\ldots,I\}) \mid j=1,\ldots J\})$. COUNT(), MIN(), MAX(), SUM() are all distributive. In fact, *F* = *G* for all but COUNT(). G= SUM() for the COUNT() function. Once order is imposed, the cumulative aggregate functions also fit in the distributive class.

**Algebraic**: Aggregate function *F()* is algebraic if there is an *M*-tuple valued function *G()* and a function *H()* such that $F(\{X_{i,j}\}) = H(\{G(\{X_{i,j} \mid i=1,..., I\}) \mid j=1,..., J\})$. Average(), standard deviation, MaxN(), MinN(), center_of_mass() are all algebraic. For Average, the function *G()* records the sum and count of the subset. The *H()* function adds these two components and then divides to produce the global average. Similar techniques apply to finding the *N* largest values, the center of mass of group of objects, and other algebraic functions. The key to algebraic functions is that a fixed size result (an M-tuple) can summarize the sub-aggregation.

**Holistic**: Aggregate function *F()* is holistic if there is no constant bound on the size of the storage needed to describe a sub-aggregate. That is, there is no constant *M*, such that an *M*-tuple characterizes the computation $F(\{X_{i,j} \mid i=1,...,I\})$. Median(), MostFrequent() (also called the Mode()), and Rank() are common examples of holistic functions.

We know of no more efficient way of computing super-aggregates of holistic functions than the $2^N$-algorithm using the standard `GROUP BY` techniques. We will not say more about cubes of holistic functions.

Cubes of distributive functions are relatively easy to compute. Given that the core is represented as an *N*-dimensional array in memory, each dimension having size $C_i+1$, the *N-1* dimensional slabs can be computed by projecting (aggregating) one dimension of the core. For example the following computation aggregates the first dimension.

`CUBE`(`ALL`, $x_2$,...,$x_N$) = F({`CUBE`(i, $x_2$,...,$x_N$) | i = 1,...$C_1$}).

*N* such computations compute the *N-1* dimensional super-aggregates. The distributive nature of the function F() allows aggregates to be aggregated. The next step is to compute the next lower dimension -- an (...`ALL`,..., `ALL`...) case. Thinking in terms of the cross tab, one has a choice of computing the result by aggregating the lower row, or aggregating the right column (aggregate (`ALL`, *) or (*, `ALL`)). Either approach will give the same answer. The algorithm will be most efficient if it aggregates the smaller of the two (pick the * with the smallest $C_i$). In this way, the super-aggregates can be computed dropping one dimension at a time.

Algebraic aggregates are more difficult to compute than distributive aggregates. Recall that an algebraic aggregate saves its computation in a handle and produces a result in the end -- at the `Final()` call. `Average()` for example maintains the count and sum values in its handle. The super-aggregate needs these intermediate results rather than just the raw sub-aggregate. An algebraic aggregate must maintain a handle (*M*-tuple) for each element of the cube (this is a standard part of the group-by operation). When the core `GROUP BY` operation completes, the `CUBE` algorithm passes the set of handles to each *N-1* dimensional super-aggregate. When this is done the handles of these super-aggregates are passed to the super-super aggregates, and so on until the (`ALL`, `ALL`, ..., `ALL`) aggregate has been computed. This approach requires a new call for distributive aggregates:

```
Iter_super(&handle, &handle)
```

which folds the sub-aggregate on the right into the super aggregate on the left. The same ordering idea (aggregate on the smallest list) applies at each higher aggregation level.

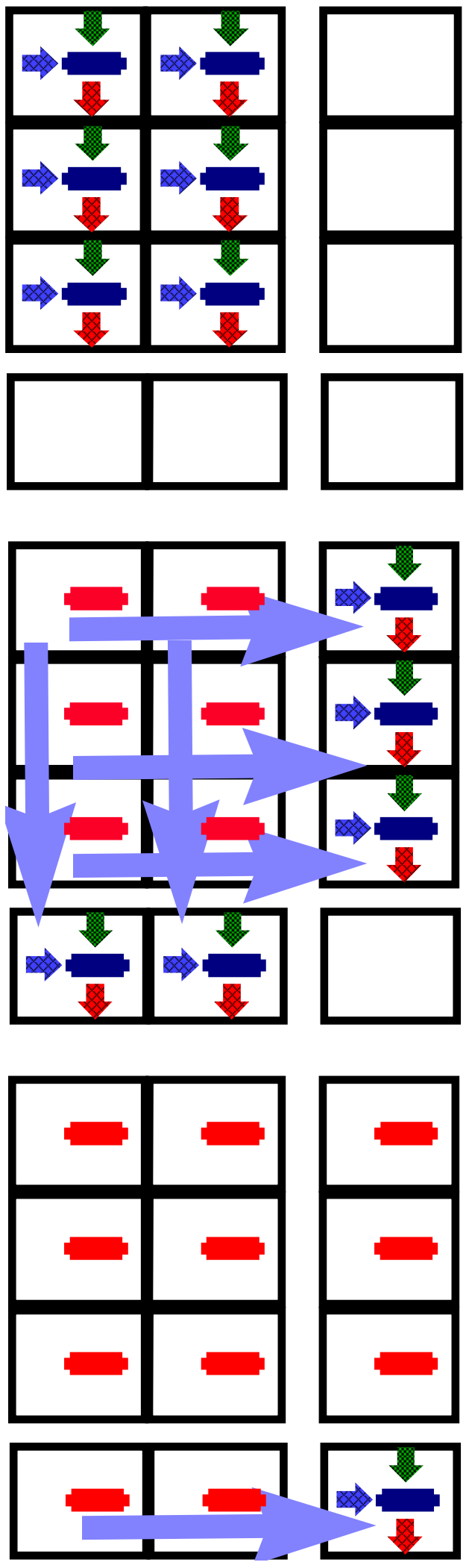

**Figure 8:** Computing the cube with a minimal number of calls to aggregation functions. If the aggregation operator is algebraic or distributive, then it is possible to compute the core of the cube as usual.

Then, the higher dimensions of the cube are computed by calling the super-itterator function passing the lower-level scratch-pads.

Once an *N*-dimensional space has been computed, the operation repeats to compute the *N-1* dimensional space. This repeats until *N=0*.

Interestingly, the distributive, algebraic, and holistic taxonomy is very useful in computing aggregates for parallel database systems. In those systems, aggregates are computed for each partition of a database in parallel. Then the results of these parallel computations are combined. The combination step is very similar to the logic and mechanism used in Figure 8.

If the data cube does not fit into memory, array techniques do not work. Rather one must either partition the cube with a hash function or sort it. These are standard techniques for computing the `GROUP BY`. The super-aggregates are likely to be orders of magnitude smaller than the core, so they are very likely to fit in memory. Sorting is especially convenient for `ROLLUP` since the user often wants the answer set in a sorted order – so the sort must be done anyway.

It is possible that the core of the cube is sparse. In that case, only the non-null elements of the core and of the super-aggregates should be represented. This suggests a hashing or a B-tree be used as the indexing scheme for aggregation values [Essbase].

## 6. Maintaining Cubes and Roll-ups

SQL Server 6.5 has supported the CUBE and ROLLUP operators for about six months now. We have been surprised that some customers use these operators to compute and store the cube. These customers then define triggers on the underlying tables so that when the tables change, the cube is dynamically updated.

This of course raises the question: how can one incrementally compute (user-defined) aggregate functions after the cube has been materialized? Harinarayn, Rajaraman, and Ullman have interesting ideas on pre-computing a sub-cubes of the cube assuming all functions are holistic [Harinarayn, Rajaraman, and Ullman]. Our view is that users avoid holistic functions by using approximation techniques. Most functions we see in practice are distributive or algebraic. For example, medians and quartiles are approximated using statistical techniques rather than being computed exactly.

The discussion of distributive, algebraic, and holistic functions in the previous section was completely focused on SELECT statements, not on UPDATE, INSERT, or DELETE statements.

Surprisingly, the issues of maintaining a cube are quite different from computing it in the first place. To give a simple example: it is easy to compute the maximum value in a cube – max is a distributive function. It is also easy to propagate inserts into a "max" N-dimensional cube. When a record is inserted into the base table, just visit the 2N super-aggregates of this record in the cube and take the max of the current and new value. This computation can be shortened -- if the new value "loses" one competition, then it will lose in all lower dimensions. Now suppose a delete or update changes the largest value in the base table. Then $2^N$ elements of the cube must be recomputed. The recomputation needs to find the global maximum. This seems to require a recomputation of the entire cube. So, max is a distributive for `SELECT` and `INSERT`, but it is holistic for `DELETE`.

This simple example suggests that there are orthogonal hierarchies for `SELECT`, `INSERT`, and `DELETE` functions (update is just delete plus insert). If a function is algebraic for insert, update, and delete (count() and sum() are such a functions), then it is easy to maintain the cube. If the function is distributive for insert, update, and delete, then by maintaining the scratchpads for each cell of the cube, it is fairly inexpensive to maintain the cube. If the function is delete-holistic (as max is) then it is expensive to maintain the cube. These ideas deserve more study.

## 7. Summary:

The cube operator generalizes and unifies several common and popular concepts:

aggregates,
group by,
histograms,
roll-ups and drill-downs and,
cross tabs.

The cube operator is based on a relational representation of aggregate data using the `ALL` value to denote the set over which each aggregation is computed. In certain cases it makes sense to restrict the cube operator to just a roll-up aggregation for drill-down reports.

The data cube is easy to compute for a wide class of functions (distributive and algebraic functions). SQL's basic set of five aggregate functions needs careful extension to include functions such as rank, N_tile, cumulative, and percent of total to ease typical data mining operations. These are easily added to SQL by supporting user-defined aggregates. These extensions require a new super-aggregate mechanism to allow efficient computation of cubes.

## 7. Acknowledgments

Joe Hellerstein suggested interpreting the `ALL` value as a set. Tanj Bennett, David Maier and Pat O'Neil made many helpful suggestions that improved the presentation.